# Materials preparation, single crystal growth, and the phase diagram of the cuprate high temperature superconductor $La_{1.6-x}Nd_{0.4}Sr_xCuO_4$

Mirela Dragomir,[1,2,3] Qianli Ma,[4] J. Patrick Clancy,[4] Amirreza Ataei,[5] Paul A. Dube,[2] Sudarshan Sharma,[4] Ashfia Huq,[6] Hanna A. Dabkowska,[2] Louis Taillefer,[5,7] and Bruce D. Gaulin[2,4,7]

[1] *Department of Chemistry and Chemical Biology, McMaster University, Hamilton, Ontario L8S 4M1, Canada*
[2] *Brockhouse Institute for Materials Research, Hamilton, Ontario L8S 4M1, Canada*
[3] *Electronic Ceramics Department, Jožef Stefan Institute, Ljubljana 1000, Slovenia*
[4] *Department of Physics and Astronomy, McMaster University, Hamilton, Ontario L8S 4M1, Canada*
[5] *Institut Quantique, Département de Physique & RQMP, Université de Sherbrooke, Sherbrooke, Québec J1K 2R1, Canada*
[6] *Neutron Scattering Division, Oak Ridge National Laboratory, Oak Ridge, Tennessee 37831, USA*
[7] *Canadian Institute for Advanced Research, Toronto, Ontario M5G 1M1, Canada*

One branch of the La-214 family of cuprate superconductors, $La_{1.6-x}Nd_{0.4}Sr_xCuO_4$ (Nd-LSCO), has been of significant and sustained interest, in large part because it displays the full complexity of the phase diagram for canonical hole-doped, high $T_C$ superconductivity, while also displaying relatively low superconducting critical temperatures. The low superconducting $T_C$'s imply that experimentally accessible magnetic fields can suppress the superconductivity to zero temperature. In particular, this has enabled various transport and thermodynamic studies of the $T = 0$ ground state in Nd-LSCO, free of superconductivity, across the critical doping $p^* = 0.23$ where the pseudogap phase ends. The strong dependence of its superconducting properties on its crystal symmetry has itself motivated careful studies of the Nd-LSCO structural phase diagram. This paper provides a systematic study and summary of the materials preparation and characterization of both single crystal and polycrystalline samples of Nd-LSCO. Single-phase polycrystalline samples with $x$ spanning the range from 0.01 to 0.40 have been synthesized, and large single crystals of $La_{1.6-x}Nd_{0.4}Sr_xCuO_4$ for select $x$ across the region (0.07, 0.12, 0.17, 0.19, 0.225, 0.24, and 0.26) were grown by the optical floating zone method. Systematic neutron and X-ray diffraction studies on these samples were performed at both low and room temperatures, 10 K and 300 K, respectively. These studies allowed us to follow the various structural phase transitions and propose an updated structural phase diagram for Nd-LSCO. In particular, we found that the low-temperature tetragonal (LTT) phase ends at a critical doping $p_{LTT} = 0.255\pm0.005$, clearly separated from $p^*$.

# I. INTRODUCTION

The discovery of high temperature superconductivity in Ba-doped $La_2CuO_4$ (LBCO) by Bednorz and Müller in 1986 [1] represented a major step forward towards achieving room-temperature superconductivity and sparked huge interest in the copper oxide-based materials referred to as "cuprates". The parent compound of LBCO is $La_2CuO_4$ (LCO) which is composed of perovskite layers of $CuO_6$ octahedra and rock salt layers of LaO structural units located above the $CuO_2$ planes. By substituting $La^{3+}$ with a non-isovalent cation, such as $Ba^{2+}$ or $Sr^{2+}$, holes are introduced into oxygen levels, and Zhang-Rice singlets formed between the oxygen and copper ions introduce "hole-doping" into the insulating state that characterizes LCO. Consequently, the electronic properties of LBCO and Sr-doped LCO (LSCO) change dramatically as a function of $Ba^{2+}$ or $Sr^{2+}$ concentration, and a rich electronic phase diagram results which features both three and two-dimensional quantum antiferromagnetism and superconductivity.

Soon after the discovery of high temperature superconductivity in LBCO, it was found that by varying the number of $CuO_2$ layers (defined by the Cu $3d_{x^2-y^2}$ orbital bonding to the O $2p_\sigma$ orbitals), the superconducting $T_C$'s of the cuprates could be greatly increased [2,3]. In this way, record superconducting $T_C$'s were achieved in the bilayer cuprate systems $Bi_2Sr_2Ca_2Cu_3O_{10}$ [3] and $Bi_2Sr_2CaCu_2O_8$ (BSCCO or Bi-2212) [4] and $YBa_2Cu_3O_{7-x}$ (YBCO or 123) [5] with $T_C$ approaching 100 K. Clearly, the superconductivity in these systems originates from the $CuO_2$ planes, but the mechanism underlying the superconductivity is enigmatic, and has remained so more than 30 years after its original discovery. Thereafter, a myriad of studies was carried out, fuelled by the need to understand the relationships among the collective states characterizing this rich phase diagram; that is between the magnetism, charge ordering, metallic and superconducting states, as well as the crystallographic structures that these materials exhibit.

The structures of the single-layer cuprates, LBCO and LSCO, are relatively simple compared with those adopted by the multi-layered compounds, which, among other things, can exhibit incommensurate structures [6]. From this point of view, the LCO-based cuprates are more amenable to a comprehensive characterization of their structures and phase diagrams.

The structures adopted by LCO and hole-doped LCO are presented in **Figure 1 a–e**. LSCO undergoes a second order phase transition from a high temperature tetragonal structure with the *I*4/*mmm* space group (HTT) illustrated in **Figure 1a**, to a low temperature orthorhombic structure, LTO, with the *Bmab* space group, illustrated in **Figure 1b**. This transition is relatively subtle and involves a cooperative tilting of the $CuO_6$ octahedra along the [110] axes of the HTT structure [7] and an elastic deformation of their basal planes, as shown in **Figure 1c**. As a consequence, the space group changes and the unit cell of the LTO phase is rotated by 45° and expanded by $\sqrt{2}$, such that new superlattice reflections appear whose intensity is proportional to the square of the tilt angle. Therefore, this tilting is the primary order parameter of the transition. In addition, splitting of the main reflections arises from the twinned orthorhombic domains that are present.

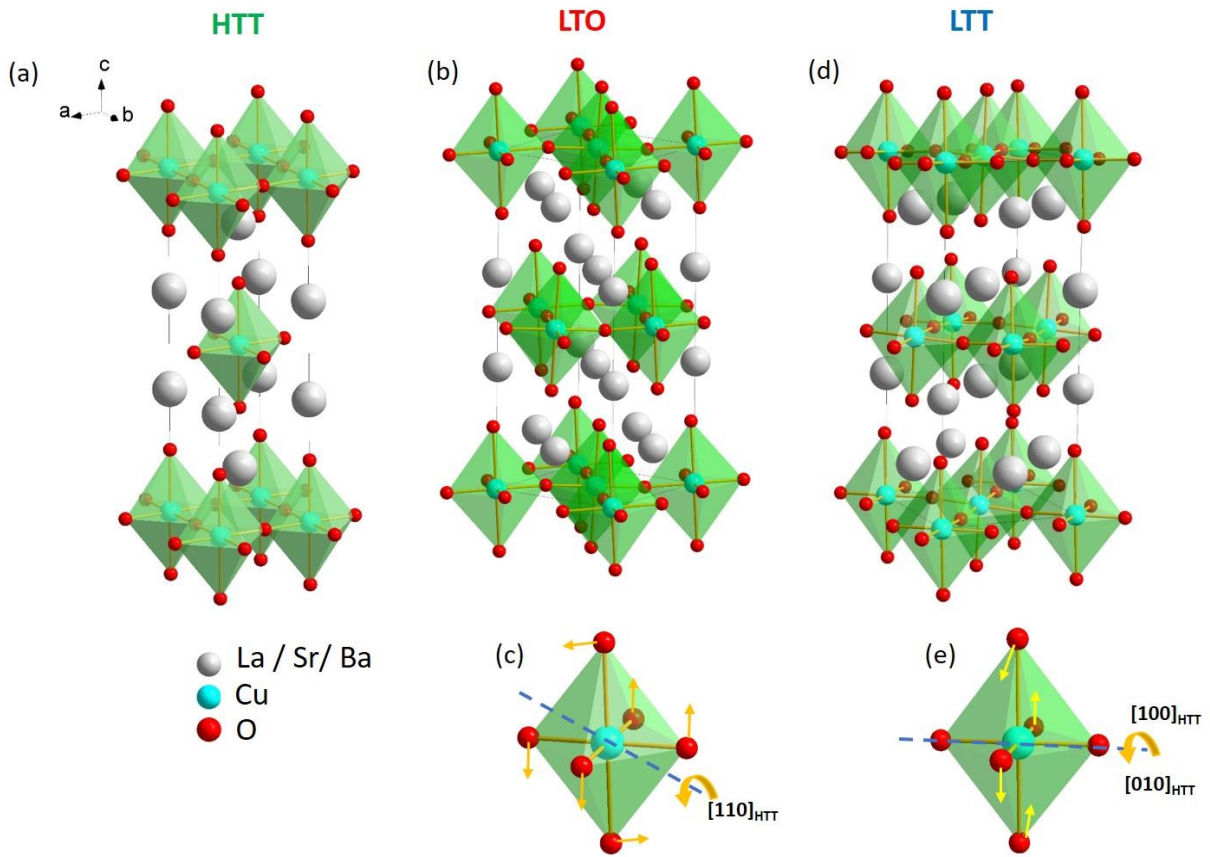

**Figure 1** Schematic views of the crystal structures of LCO and hole-doped LCO: **(a)** high temperature tetragonal, HTT; **(b)** low temperature orthorhombic, LTO; **(c)** The rotations or tilting of the $CuO_6$ octahedra within the LTO phase relative to the HTT phase; **(d)** low-temperature tetragonal, LTT, phase; **(e)** The rotations or tilting of the $CuO_6$ octahedra within the LTT phase relative to the HTT phase .

The hole-doped LCO structure in LBCO or LSCO changes with both temperature and hole doping concentration, $x$ [8]. For a given hole doping, and as a function of decreasing temperature, a sequence of structural transitions occurs in LCO materials: $T$ (HTT) > $T$ (LTO) > $T$ (LTT), where HTT represents the tetragonal, high temperature non-tilted structure, and LTO and LTT are the two, low temperature, tilted structures [7,9]. Although the LTT designation stands for "Low Temperature Tetragonal", this structure is in fact orthorhombic with space group ($P4_2/ncmz$), and is illustrated in **Figure 1d**. The LTO to LTT transition is first order (discontinuous) in nature and involves tilts about the [100] axes of the HTT structure - **Figure 1e**. The LTT structure is therefore not a subgroup of the LTO structure, which displays tilts about [110] relative to the HTT structure, as illustrated in **Figure 1c**. The discontinuous nature of the structural phase transition between the LTO and LTT phases is thus expected.

The phase diagram of $La_{1.6-x}Nd_{0.4}Sr_xCuO_4$ (Nd-LSCO) is very similar to that of the other single-layer hole-doped cuprates and as a function of hole doping $p$ (= $x$), it moves from a three dimensional commensurate antiferromagnet ($x = 0$), to a region of static two-dimensional spin stripe order, to spin and charge stripes co-existing with superconductivity at low temperatures, and finally to a conventional non-superconducting Fermi liquid beyond $x \sim 0.26$ [10,11]. The maximum superconducting $T_C$ is ~ 15 K at $x \sim 0.19$. Below a crossover temperature $T^*$ there exists an enigmatic part of the phase diagram referred to as the "pseudogap phase", characterised by a gapped electronic density of states [12,13]. In Nd-LSCO, ARPES measurements reveal that an anti-nodal pseudogap is present at dopings up to $p = 0.20$, but is absent at $p = 0.24$ [14]. Very recent work [15] has shown thermodynamic evidence for a *quantum critical point* at $p^* = 0.23 \pm 0.005$ in Nd-LSCO, associated with the end of this line of phase transitions ($T^*(p)$) [16]. The relatively low maximum $T_C$ exhibited by the Nd-LSCO system was crucial for this study, as it allowed the suppression of superconductivity by practical magnetic fields ($H \leq 18$ T), so that the thermodynamic properties of the normal state at these doping levels, $p$, could be studied. The electronic heat capacity of the Nd-LSCO system could then be measured over a range of hole concentrations spanning the underdoped, optimally-doped, and overdoped regions, and as a function of temperature. Michon *et al*. [15] found that the $C_{el}/T$ is strongly peaked at $p^*$ and that at $p^*$ it follows a $\log(1/T)$ dependence as $T$ tends to zero. These findings are typical thermodynamic signatures of a quantum critical point, as previously observed, for example, in iron pnictide superconductors [17] and heavy-fermion metals [18].

In Nd-LSCO, a number of transport properties are seen to change abruptly across $p^* = 0.23$, in the $T = 0$ limit and without superconductivity. The carrier density drops from $n \sim 1+p$ above $p^*$ to $n \sim p$ below $p^*$, as measured by the Hall coefficient [19,20,21], the electrical resistivity [19,20] and the thermal conductivity [22]. A new contribution to the thermal Hall conductivity from neutral excitations appears below $p^*$ [23], and the Fermi surface changes topology below $p^*$, as seen by angle-dependent magneto-resistance [24].

According to the structural phase diagram of Nd-LSCO first reported in the literature by Axe and Crawford [9], as well as in other studies [2,3,25], the following structural hierarchy is observed at constant hole-concentration for $x \leq 0.15$: $T$ (HTT) > $T$ (LTO1) > $T$ (LTO2) > $T$ (LTT). Note that relative to both LBCO and LSCO, an additional low temperature orthorhombic phase, LTO2 (*Pccn* space group), is observed in the Nd-LSCO system, which seems to separate the LTO1 and LTT structural phases at low hole doping concentrations [26,27]. The LTO2 structure is obtained by simultaneous and unequal tilts about the [110] and [1−10] axes of the HTT structure [28].

Among the key questions that ensue then, are what role do the structural symmetries of these crystalline phases play on the superconducting and pseudogap properties of Nd-LSCO? The pre-existing phase diagram for Nd-LSCO covers only Sr concentrations up to $x = 0.25$, where superconductivity seems to end, but information near these concentrations and at higher doping is lacking. This is likely due to the difficulty associated with synthesising samples with $x \geq 0.2$ including overdoped samples of this and other high temperature $T_C$ superconductors [29].

In this study, we report on the synthesis and characterisation of both polycrystalline and large single crystals of Nd-LSCO, appropriate for neutron scattering studies, for example. Our group has synthesized single-phase, polycrystalline Nd-LSCO samples with $x$ spanning from 0.01 to 0.40, and have grown large single crystals (~ 5 grams and larger) of $La_{1.6-x}Nd_{0.4}Sr_xCuO_4$ for select nominal $x = 0.07, 0.12, 0.17, 0.19, 0.225, 0.24$, and $0.26$. This work thus extends our understanding of the structural phase diagram of Nd-LSCO at both low and high hole-doping, and in particular provides greater detail in the regime around the end of the pseudogap phase, at $p^* = 0.23$, and the end of superconductivity, at $p_c \sim 0.27$. We describe the synthesis of single-phase powder samples as well as the single crystals, and report systematic neutron and X-ray diffraction studies characterizing these samples at both low and high temperatures. These analyses allow us to ascertain the stoichiometries of the samples we have synthesized, and benchmark them against the results of earlier studies. These new samples have been and will be, the subject of further experimental studies. They also allow us to propose an updated phase diagram for Nd-LSCO that now covers Sr doping concentrations up to $x = 0.40$, and make comparisons to the corresponding phase diagram for LSCO.

## II. EXPERIMENTAL SECTION

## A. Materials preparation

### 1. Polycrystalline samples

The chemical reaction that led to the formation of the $La_{1.6-x}Nd_{0.4}Sr_xCuO_4$ ($x$ = 0.01, 0.02, 0.04, 0.05, 0.07, 0.12, 0.17, 0.19, 0.225, 0.24, 0.26, 0.27, 0.36, and 0.40) compounds can be written as:

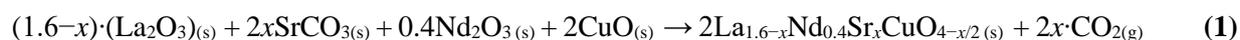

$(1.6-x)\cdot(La_2O_3)_{(s)} + 2xSrCO_{3(s)} + 0.4Nd_2O_{3\,(s)} + 2CuO_{(s)} \rightarrow 2La_{1.6-x}Nd_{0.4}Sr_xCuO_{4-x/2\,(s)} + 2x\cdot CO_{2(g)}$ **(1)**

The starting materials used in the synthesis of $La_{1.6-x}Nd_{0.4}Sr_xCuO_4$ were: $La_2O_3$ (Cerac, 99.99%), $Nd_2O_3$ (REO, Inframat Adv. Mat., 99.9%), CuO (Cerac, 99.999%), and $SrCO_3$ (Cerac, 99.994%). The lanthanides, $La_2O_3$ and $Nd_2O_3$, were first pre-annealed at 1000 °C to remove any residual water. An appropriate stoichiometric mixture of the oxides was homogenised in a planetary mill, dried, followed by homogenisation by hand, in an agate mortar, for about 1 h. The as-obtained homogenized powder was annealed several times at 1000 °C, in air, with intermittent grinding. Samples with higher Sr concentration, $x$, required longer annealing times of up to 1 week. The resulting powder samples were of black colour with a greyish shine.

### 2. Single crystal growth

Polycrystalline $La_{1.6-x}Nd_{0.4}Sr_xCuO_4$ with $x$ = 0.07, 0.12, 0.17, 0.19, 0.225, 0.24, and 0.26 were pressed into rods ~80 mm long and with a diameter of 8 mm. A stable single crystal growth requires dense, single-phase rods. For this purpose, the polycrystalline rods were then annealed for 48 h at temperatures around 1200 °C (the temperature slightly varied and was adjusted for each composition), under an oxygen flow of ~15 mL/min. The heating and cooling rates were 100 degrees per hour. To compensate for the loss of CuO during annealing and crystal growth, an excess of 3 at. % CuO was used.

The Nd-LSCO materials melt incongruently which implies that instead of producing a melt with the same composition as the prepared polycrystalline material, the melt has a different composition from the solid material feeding it. Therefore, in order to be able to grow large single crystals of these incongruently melting materials, the traveling solvent floating zone (TSFZ) technique needs to be employed. In contrast to congruent bulk crystal growth techniques, the TSFZ technique involves dissolving a small portion of the material into the created solvent. The solvent is designed in such a way that it melts well below the melting point of the crystal and creates a liquid reservoir. During the crystal growth, the feed rod continues to supply the reservoir with the desired composition of the target crystal and eventually induces saturation within the solvent. Finding the right solvent is never an easy task and to determine the appropriate solvent for Nd-LSCO, the phase diagram of the $La_2O_3-CuO$ binary system was considered [30]. After many different trials, an ideal ratio of 22:78 for $La_2O_3$ vs CuO was chosen as the solvent. However, adding Sr and Nd to the solvent with the ratio of the elements being similar to what they are in the desired crystal was also found to be important for a stable crystal growth.

The single crystal growths were performed on a four-mirror optical floating zone (OFZ) furnace (Crystal Systems, Inc.). The growth of Nd-LSCO single crystals has proven to be challenging due to the relatively low growth rate and their high sensitivity to the growth conditions. To maintain a stable growth, a steady length of the molten zone is desirable and it requires careful control. This can be achieved by maintaining a constant growth speed. For this system, the growth speed should be less than 1 mm/hr, as faster growth speeds will dramatically affect the crystal quality. To achieve relatively large single crystals, we determined 0.65 mm/hr as an optimal growth speed for Nd-LSCO using the four-mirror floating zone image furnace at McMaster University. During the growth, both the feed and the seed rods were coaxially counter-rotated at a rate of 30 rpm.

The resulting single crystals, which were typically 8 mm in diameter and about 30 up to 60 mm long, were used in several different experiments, but the large size of the single crystals was driven by the needs of neutron experiments. The crystals were grown in air, and were later annealed in flowing oxygen, as described in the next sections.

## B. Characterisation

### 1. X-ray powder diffraction

Laboratory X-ray powder diffraction analysis was carried out on a zero-background plate, at room temperature, using a PANalytical X'Pert PRO (PANalytical, Almelo, Netherlands) with Cu-K$\alpha_1$ radiation ($\lambda = 1.54056$ Å) and X'Celerator detector. A 2θ step of 0.0167° was used for data collection in the 15–80º 2$\theta$ range. The phase identification was performed using the PDF-2 database and the HighScore software.

### 2. Neutron powder diffraction

Time of flight (TOF) powder neutron diffraction experiments were collected on the powder Nd-LSCO samples using the POWGEN instrument [31] (BL-11A) at the Spallation Neutron Source (SNS), Oak Ridge National Laboratory, Oak Ridge, Tennessee, USA. Two sets of measurements were performed at temperatures of 300 K and 10 K using neutron beams with as wavelength centred on 1.5 Å and covering a d-spacing range of 0.5–13 Å. Roughly 500 mg of each sample was loaded in a vanadium can (6 mm inner diameter) for the measurements. The Rietveld refinements of the acquired data were performed with the GSAS-II program [32].

### 3. Single crystal X-ray diffraction

White beam (Laue) X-ray diffraction was performed on all of our single crystal samples to assess the surface quality of the as-grown crystals and to determine the orientation of the single crystal pieces. The Laue measurements were performed using a tungsten tube source powered by a Spellman power supply at a voltage of 10 kV and current of 10 mA. The diffracted X-rays were detected with a Multiwire Laboratories detector, while the data was collected with North Star software. The orientation of the crystals was determined by fitting the collected patterns using Orient Express.

High resolution single crystal X-ray scattering measurements were performed using Cu K$\alpha$ radiation ($\lambda$ = 1.54 Å) produced by an 18 kW rotating anode source with a pyrolytic graphite monochromator [PG-(0,0,2)]. Samples with a size of ~3–5 mm (cross section) and ~0.5–1 mm (thickness) were mounted on the cold finger of a closed cycle cryostat, and were aligned using a Huber 4-Circle diffractometer. The accessible temperature range for the cryostat was from $T$ = 4 K up to $T$ = 310 K. The Nd-LSCO single crystals were characterized by measuring the temperature dependence of several Bragg peaks which differ appreciably in each of the HTT, LTO, and LTT crystallographic phases. For example, the high temperature tetragonal (HTT) to low temperature orthorhombic (LTO) structural phase transition is relatively easily identified from the splitting of $(H,H,0)_{HTT}$ type peaks into $(2H,0,0)_{LTO}$ and $(0,2H,0)_{LTO}$ peaks. Similarly, the phase transitions from LTO to low temperature tetragonal (LTT) or to low temperature less orthorhombic (LTO2) phases can be identified based on the appearance of new superlattice Bragg peaks at the $(H,0,0)_{HTT}$, H = odd or $(0,K,0)_{HTT}$, K = odd positions.

### 4. Elastic neutron scattering measurements on single crystals

Elastic neutron scattering measurements were performed on large single crystals as part of a study of their full inelastic neutron scattering spectra. The results from the elastic measurements are very useful in their own right, as they permit a comprehensive check of the single crystal nature of these relatively large samples. As neutrons are deeply penetrating, such measurements are sensitive to single crystal grains throughout the material. These measurements were performed using the SEQUOIA Time of Flight Chopper Spectrometer at the Spallation Neutron Source (SNS) at Oak Ridge National Laboratory (ORNL) [33]. These measurements were carried out on large $x$ = 0.12, 0.19, and 0.24 single crystals of Nd-LSCO, and were displayed and analysed using Mantid software [34].

### 5. Magnetic susceptibility and resistivity measurements

Magnetic susceptibility measurements were carried out on single crystals of Nd-LSCO using a Quantum design MPMS SQUID magnetometer. The direct current zero-field cooled and field-cooled data were collected in the temperature range from 2 to 30 K and an applied magnetic field ranging from 10 Oe to 1000 Oe. The single crystals were mounted on a quartz sample holder and the magnetic fields were applied along the $c$-axis.

Samples for resistivity measurements were cut in the shape of rectangular platelets with a cross-section size of typically 2 mm$^2$ and thickness of 0.2 mm. The $a$ and $c$ axis of the samples were along its largest and smallest dimensions. The longitudinal and transverse current pairs of contacts were made using silver paste epoxy H20E EpoTek and silver wires with a diameter of 0.025 mm. The contacts were diffused in the sample by annealing in an oxygen atmosphere at 500 °C for 1 hour. The resistivity was measured longitudinally with the standard four-probe technique. In all the measurements, the current was applied along the length of the samples.

# III. RESULTS AND DISCUSSION

## A. Polycrystalline samples

### 1. NPD and XRD at 300 K

Examination of the Nd-LSCO powder samples performed with our lab-based CuK$\alpha_1$ X-ray source at room temperature confirmed the phase purity and crystalline nature of the synthesised powders. Room-temperature neutron diffraction measurements were also carried out. Both room-temperature X-ray and neutron powder diffraction measurements were analysed in a low temperature orthorhombic type structure (LTO1), space group *Bmab*, for $x \leq 0.12$, and a high-temperature tetragonal structure (HTT), space group *I4mmm*, for $x > 0.12$.

**Figure 2** presents the neutron powder diffraction profiles obtained at room temperature for the entire powder series, $0.01 \leq x \leq 0.40$. At 300 K, the evolution of the phase transition from the HTT (*I4mmm*) to LTO1 (*Bmab*) in the powder samples can be observed by following the splitting of the $(110)_{HTT}$ reflection into the (020) and (200) reflections appropriate to *Bmab* symmetry, as $x$ varies from 0.02 to 0.4, as shown in **Figure 2a**. The primary difference between the LTO and HTT structures is the tilt of the CuO$_6$ octahedra about the [110] axes of the HTT structure. Due to this tilt, the unit cell rotates by 45° and increases by $\sqrt{2}$ and new superlattice reflections appear with intensities proportional to square of the tilt angle, which increases with lowering the temperature, but decreases with increasing $x$. Therefore, this phase transition can also be followed by monitoring the evolution of the superlattice reflection as shown in **Figure 2b**. Here, we can see that there is a continuous decrease in the intensity of this reflection as the orthorhombicity decreases, characteristic of a continuous, second order phase transition.

By integrating the relative intensity of the superlattice reflection as a function of Sr composition, $x$, as shown in **Figure 2c**, we can accurately estimate the critical concentration for which the LTO structure exists to be $x = 0.16 \pm 0.01$. These results are in good agreement with the previous report by Axe *et al.* [9] which covered compositions in powder samples from $x = 0.05$ up to 0.25.

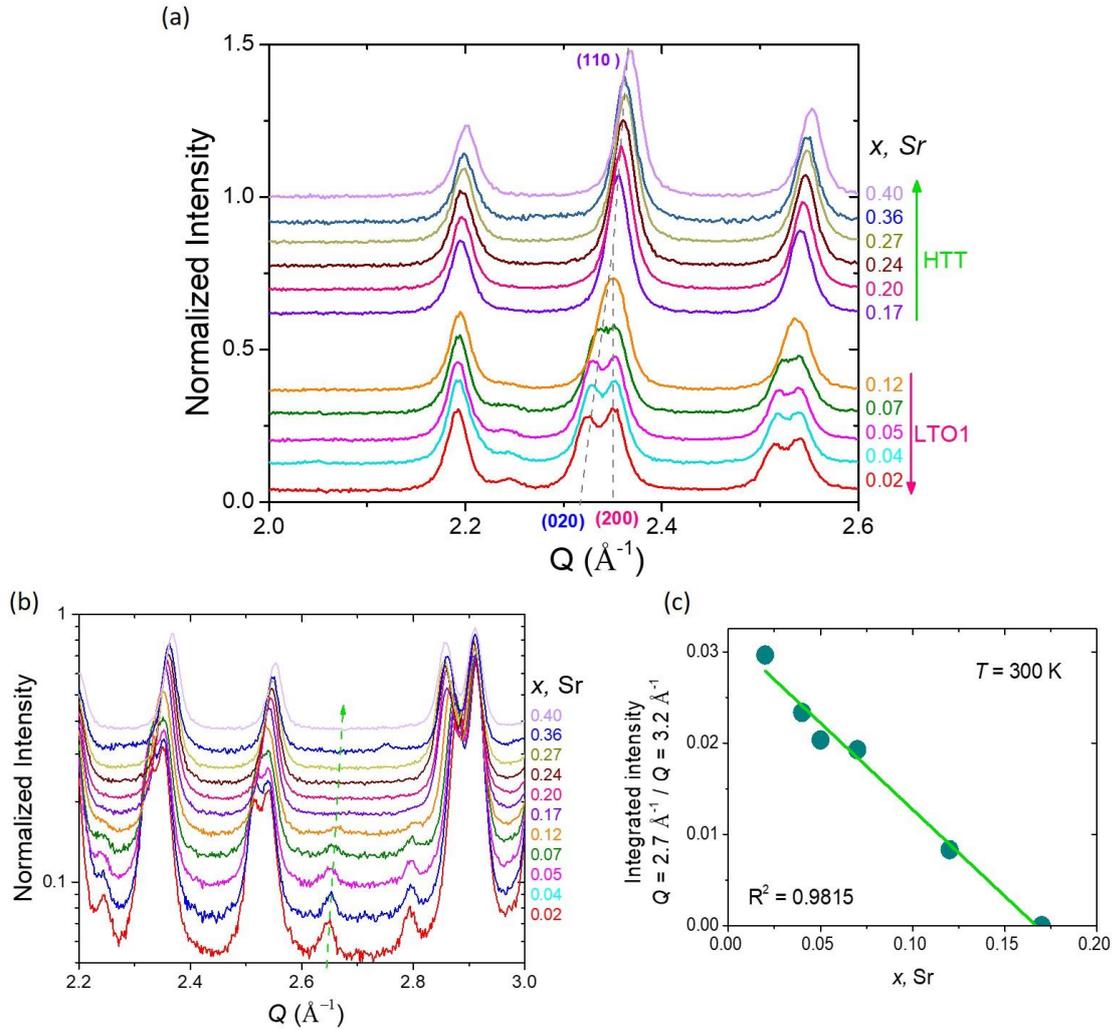

**Figure 2 (a)** The evolution of the neutron diffraction profiles for Nd-LSCO as a function of *x*, at room temperature, *T* = 300 K. The phase transition from HTT to LTO (or LTO1) can be followed either by the splitting of the (110) reflection to (020) and (200) as in **(a)**, or by the evolution of the intensities of the superlattice reflections as a function of *x*, which is more easily seen on the logaritmic intensity scale **(b)**. **(c)** The plot of the relative intensity of the superlattice intensity as a function of *x*, showing the disappearence of the superlatice intensity at $x = 0.16 \pm 0.01$.

A plot comparing the refined lattice parameters vs composition for Nd-LSCO at 300 K, for both X-ray and neutron diffraction data, is presented in **Figure 3.** For comparison purposes, $\sqrt{a}$ is plotted for the tetragonal cell. Nearly identical refined lattice parameters were obtained from the refinement of these two sets of data (X-ray and neutron). The *c* lattice parameter slowly increases with *x* while the *b* lattice parameter shows a large initial decrease with *x*. No anomalous behaviour in these parameters as a function of *x* is observed, giving further confidence that these polycrystalline samples are single phase and thus confirming the stoichiometry of our samples. The evolution of the *a*, *b*, and *c* lattice parameters in Nd-LSCO as a function of *x*, resulting from the Rietveld refinement of the neutron powder diffraction data at 300 K and compared to the data at 10 K are plotted in **Figure 3c**.

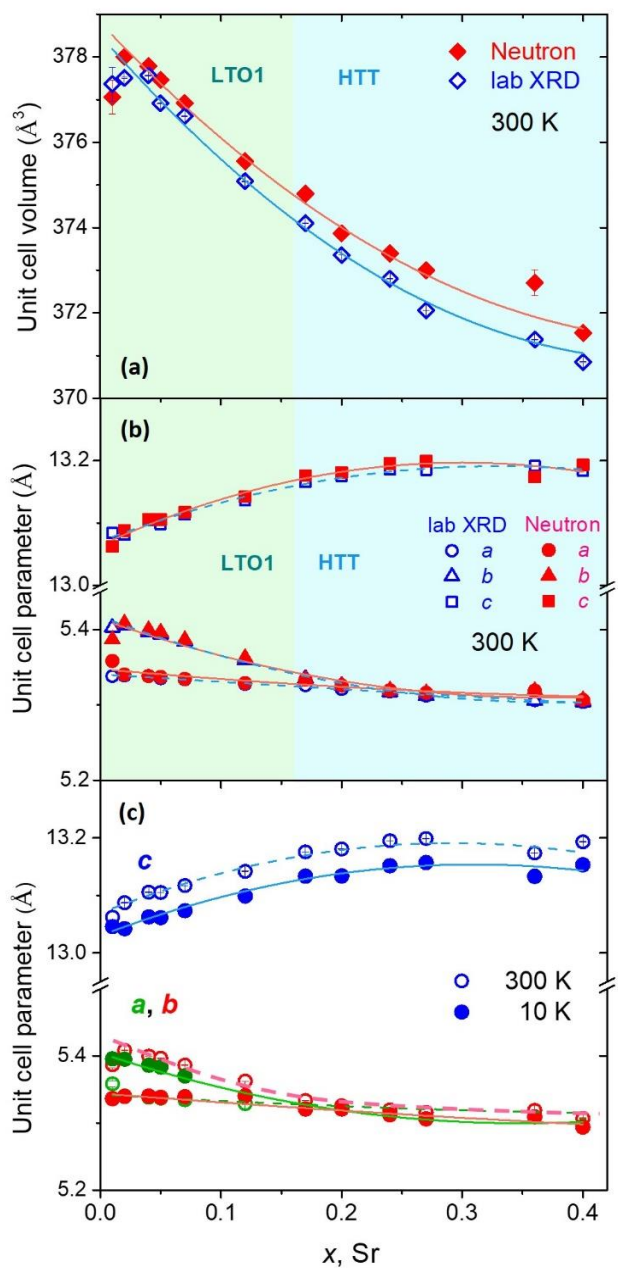

**Figure 3** The refined unit cell volume, **(a),** and lattice parameters, **(b),** vs. composition for Nd-LSCO at 300 K with the *Bmab* and *I*4/*mmm* space groups, obtained from Rietveld refinement analysis of both X-ray and TOF-NPD data. $\sqrt{a}$ is plotted for those samples displaying HTT structures. **(c)** Lattice parameters, *a*, *b* and *c* of Nd-LSCO at 300 and 10 K as function of Sr concentration, *x*, as obtained from the refinement of the neutron powder diffraction data.

The building blocks of the $La_{1.6-x}Nd_{0.4}Sr_xCuO_4$ structure are the $CuO_6$ octahedra where the Cu atoms are bonded with 4 planar oxygen atoms (O1) and 2 apical oxygen atoms (O2). Due to the Jahn-Teller effect, the octahedra are distorted and there is a strong overlap of the Cu−O orbitals. The evolution of the Cu−O bond lengths as a function of $x$ at $T = 300$ K is plotted in **Figure 4**. It can be seen that the apical Cu−O bond increases up to $x = 0.07$ after which it experiences a slow decrease, while the planar Cu−O bond lengths decrease almost linearly with $x$. The evolution of these bond lengths is related not only with the structural changes, but also but also with the increase in the hole concentration.

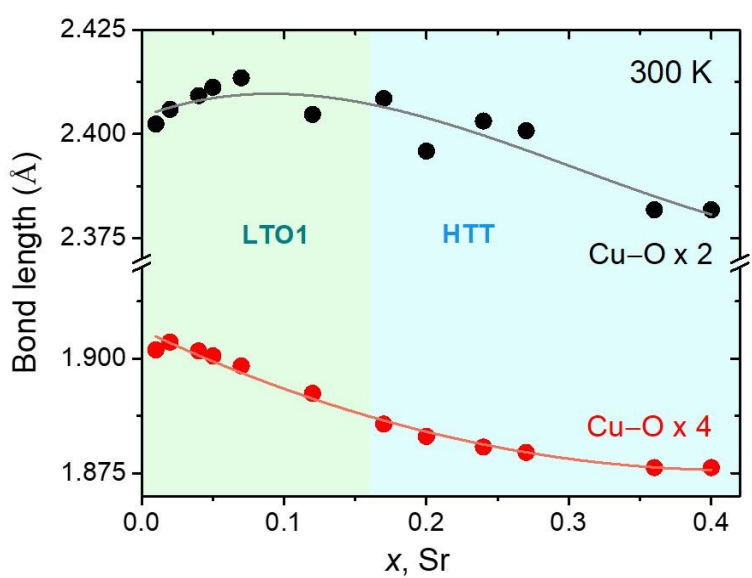

**Figure 4** Cu−O bond lengths for both planar (Cu−O x 4) and apical (Cu−O x 2) oxygen ions of $La_{1.6-x}Nd_{0.4}Sr_xCuO_4$ as a function of $x$, obtained from Rietveld refinement of the NPD data at 300 K.

**Figure 5** shows representative examples of the Rietveld neutron refinement profiles for the powder samples with $x = 0.12$ and $0.40$, while **Table I** summarizes the structural parameters resulting from these room temperature refinements. The room-temperature structure for $x = 0.12$ is orthorhombic with the *Bmab* space group. During the refinement, other structural models, such as *P4₂/ncmz* and *Pccn,* were also considered to fit the experimental neutron powder diffraction data, but were ruled out by the indexing and refinement analysis. The structure of the sample with $x = 0.40$ easily refines in the *I4/mmm* space group.

These refinements of the power diffraction data show that the quality of the diffraction patterns and their refinements are very similar across the full spectrum of hole-doping studied, from $x = 0.01$ to $x = 0.4$. In addition, we can estimate the uncertainty associated with the stoichiometries of the synthesized materials on the basis of these refinements. We see that the oxygen occupations are stoichiometric at the $\pm 2\%$ level, while the Sr and Nd concentrations are appropriate to the target for the synthesis at the $\pm 1\%$ level or better.

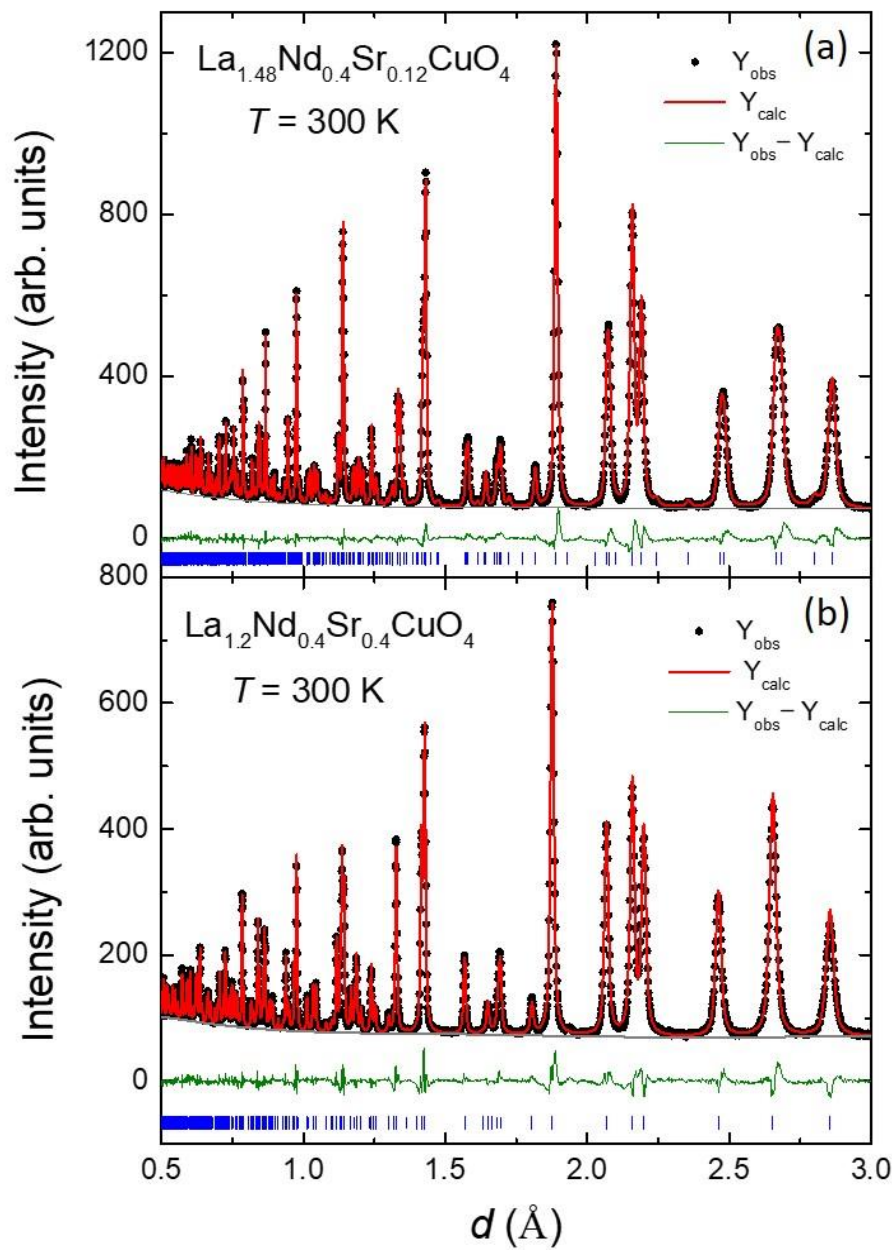

**Figure 5** Rietveld refinement profiles of the neutron diffraction data obtained at room temperature for $x = 0.12$ **(a)** and $x = 0.40$ **(b)**. The measured pattern is shown in black, the calculated profile in red, difference in green, while thick marks represent the allowed Bragg peaks.

**Table I** Results of the Rietveld neutron refinement for Nd-LSCO with $x = 0.12$ and $0.40$, $T = 300$ K: lattice parameters, positional, and thermal parameters are displayed together with the tilt angle.

| $x = 0.12$ | | | | | | | |
|---|---|---|---|---|---|---|---|
| Atom | Space group | Wycoff site | $x$ | $y$ | $z$ | $U_{iso}$ | Occ |
| La | | 8f | 0 | 0.0043(3) | 0.3606(1) | 0.005(3) | 0.742(1) |
| Nd | | 8f | 0 | 0.0043(3) | 0.3606(1) | 0.006(2) | 0.202(2) |
| Sr | | 8f | 0 | 0.0043(3) | 0.3606(1) | 0.006(1) | 0.059(2) |
| Cu | Bamb | 4a | 0 | 0 | 0 | 0.003(2) | 0.984(4) |
| O1 | | 8e | 0.25 | 0.25 | 0.0073(1) | 0.008(2) | 0.977(4) |
| O2 | | 8f | 0 | −0.0304(3) | 0.1825(1) | 0.0146(1) | 0.988(4) |
| Tilt angle | 5.7° | | | | | | |
| $R_{wp}$ | 2.97% | | | | | | |
| $a$ (Å) | 5.3289(7) | | | | | | |
| $b$ (Å) | 5.3625(8) | | | | | | |
| $c$ (Å) | 13.1421(2) | | | | | | |
| $V$ (Å$^3$) | 375.557(6) | | | | | | |
| $x = 0.40$ | | | | | | | |
| Atom | Space group | Wycoff site | $x$ | $y$ | $z$ | $U_{iso}$ | Occ |
| La | | 4e | 0 | 0 | 0.3593(1) | 0.018(5) | 0.589(7) |
| Nd | | 4e | 0 | 0 | 0.3593(1) | 0.030(5) | 0.221(7) |
| Sr | | 4e | 0 | 0 | 0.3593(1) | 0.024(5) | 0.190(3) |
| Cu | I4/mmm | 2a | 0 | 0 | 0 | 0.003(1) | 0.969(6) |
| O1 | | 4c | 0 | 0.5 | 0 | 0.009(1) | 0.968(5) |
| O2 | | 4e | 0 | 0 | 0.1804(2) | 0.017(1) | 0.980(8) |
| Tilt angle | 0° | | | | | | |
| $R_{wp}$ | 3.06% | | | | | | |
| $a$ (Å) | 3.7523(1) | | | | | | |
| $b$ (Å) | 3.7523(1) | | | | | | |
| $c$ (Å) | 13.1932(4) | | | | | | |
| $V$ (Å$^3$) | 185.761(1) | | | | | | |

### 2. Neutron powder diffraction at 10 K

Our $T = 10$ K neutron powder diffraction data reveals that the Nd-LSCO phase diagram becomes more complex at low temperatures, consistent with earlier studies, encountering three structural phase transitions with increasing Sr concentration, $x$, as shown in **Figures 6a and b**. This data was analysed using the LTO2 – *Bmab* orthorhombic space group for $0.01 \leq x \leq 0.07$, the LTT – low temperature tetragonal space group, $P4_2/ncmz$ for $0.12 \leq x \leq 0.24$, and the high temperature tetragonal, HTT – *I4/mmm* space group for $0.27 \leq x \leq 0.40$.

The CuO$_6$ octahedra tilt about different directions within the HTT structure, with tilts about the [110] direction for the LTO structure and about the [100] and [010] for the LTT structure, respectively. Studies have shown that the tilts displayed by the LTT structure play an important role in stabilizing spin and charge stripe order [35]. This induces $C_4$ intra-unit cell symmetry breaking which stabilizes stripes whose orientation rotates by 90° between neighbouring CuO$_2$ planes.

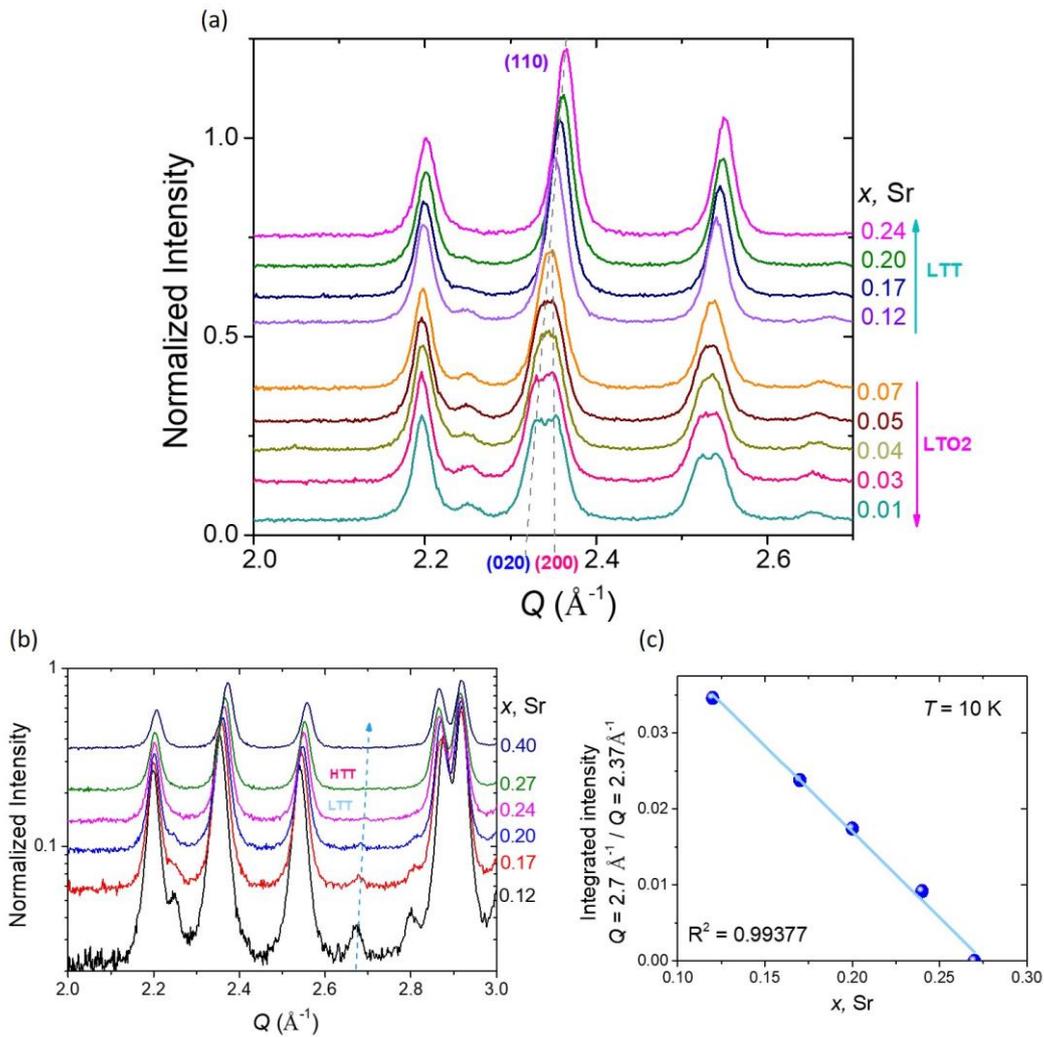

**Figure 6 (a)** The evolution of the neutron diffraction profiles for Nd-LSCO as a function of $x$, at 10 K. The phase transition from LTT to HTT was followed by the evolution of the superlattice reflection intensity as a function of $x$, which is more easily observed with intensity on a logaritmic scale **(b)**. The plot of the relative intensity of the superlattice reflection as a function of $x$ shows the disappearence of the superlatice reflection between $x = 0.24$ and $x = 0.27$. **(c)** A plot of the integrated intensity extrapolates the critical concentration for the end of the LTT phase to be $x_{LTT} \sim 0.27$.

The evolution of the Cu−O bond lengths as a function of $x$, at 10 K is shown in **Figure 7**. Note that for the LTT structure, there are two sets of Cu−O planar bonds due to the tilting of the $CuO_6$ octahedra within the LTT phase, although the difference between them is very small, only 0.28 %. With increasing Sr concentration from $x = 0.01$ to 0.40, and consequently increasing the hole doping, both the planar and apical Cu−O bonds experience a decrease, the latter ones decreasing from 2.406(2) to 2.375(2) Å. It seems that there is no compression of the $CuO_6$ octahedra with $x$ since the Cu−O apical bonds do not experience any lengthening, but rather follow the same trend as the planar bonds. The decrease in Cu−O planar bonds can be correlated with the decrease of the $a$ and $b$ lattice parameters – see **Figure 3c**. However, the decrease in the axial bonds is not correlated directly with the $c$ lattice parameter which slightly increases with $x$. This is believed to be associated with the increase of the hole concentration of the system [36]. For samples with $x > 0.27$ there is no significant change in either of the Cu−O bond lengths. We therefore conclude that the decrease in the bond lengths as $x$ increases is likely a result of the structural changes.

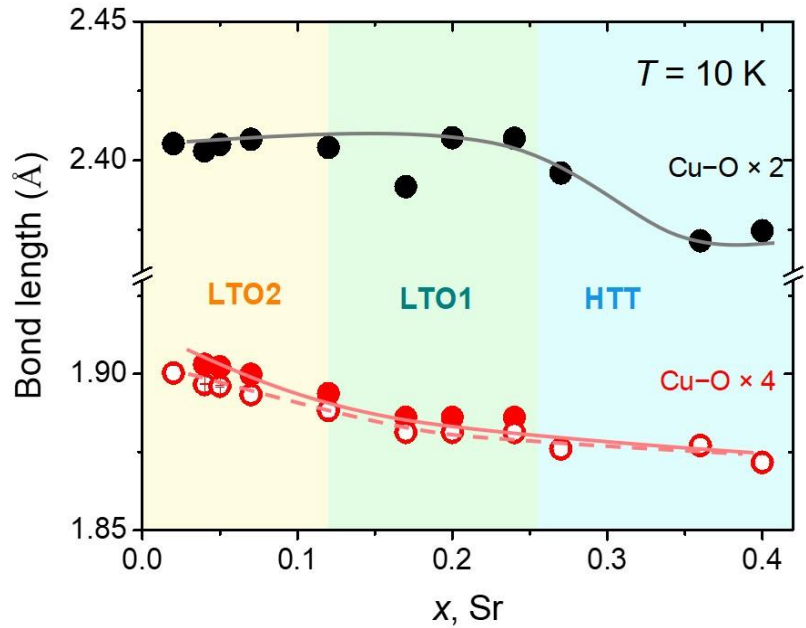

**Figure 7** The Cu−O bond lengths for both planar (2 sets of Cu−O x 2) and apical (Cu−O x 2) oxygen ions of $La_{1.6-x}Nd_{0.4}Sr_xCuO_4$ as a function of $x$, as obtained from Rietveld refinement of the neutron powder diffraction data at $T = 10$ K.

A summary of the results of the structural refinements of the Nd-LSCO polycrystalline samples using NPD (at 300 and 10 K) and XRD data (300 K) is presented in **Table II**.

**Table II** Summary of the structural refinements of the neutron powder diffraction data at 300 K and 10 K and X-ray powder diffraction data at 300 K.

| 300 K | | | 10 K | | |
|---|---|---|---|---|---|
| Sample / $x$ (Sr) | Space group | Abbreviation | Sample / $x$ (Sr) | Space group | Abbreviation |
| 0.40 | $I4/mmm$ | HTT | 0.40 | $I4/mmm$ | HTT |
| 0.36 | $I4/mmm$ | HTT | 0.36 | $I4/mmm$ | HTT |
| 0.27 | $I4/mmm$ | HTT | 0.27 | $I4/mmm$ | HTT |
| 0.24 | $I4/mmm$ | HTT | 0.24 | $P4_2/ncmz$ | LTT |
| 0.20 | $I4/mmm$ | HTT | 0.20 | $P4_2/ncmz$ | LTT |
| 0.17 | $I4/mmm$ | HTT | 0.17 | $P4_2/ncmz$ | LTT |
| 0.12 | $Bmab$ | LTO1 | 0.12 | $P4_2/ncmz$ | LTT |
| 0.07 | $Bmab$ | LTO1 | 0.07 | $Pccn$ | LTO2 |
| 0.05 | $Bmab$ | LTO1 | 0.05 | $Pccn$ | LTO2 |
| 0.04 | $Bmab$ | LTO1 | 0.04 | $Pccn$ | LTO2 |
| 0.02 | $Bmab$ | LTO1 | 0.02 | $Pccn$ | LTO2 |
| 0.01 | $Bmab$ | LTO1 | 0.01 | $Pccn$ | LTO2 |

## B. Single crystals

### *1. Single crystal characterization*

Images relevant to the growth of the Nd-LSCO single crystals where $x = 0.19$ are shown in **Figure 8**. An image of an annealed polycrystalline rod of Nd-LSCO prior to its use in a single crystal growth using a four-mirror optical floating zone furnace is shown in **Figure 8a**, while **Figure 8b** shows the single crystal resulting from the growth. To identify the single-crystalline nature and the crystal growth habit, Laue diffraction was performed perpendicular to the growth axis, on different regions of the crystal. **Figure 8c** shows a typical Laue X-ray diffraction pattern obtained. All the observed black spots were indexed assuming the growth direction was along the [001]. It can be seen that all spots were sharp within the instrumental resolution, without splitting or broadening that would indicate twinning, thus confirming its single crystal nature. Based on this analysis, we can state that the Nd-LSCO crystals grown in this study are large single grains. However, neutron scattering characterisation needs to be employed to confirm this and we do so in the next section.

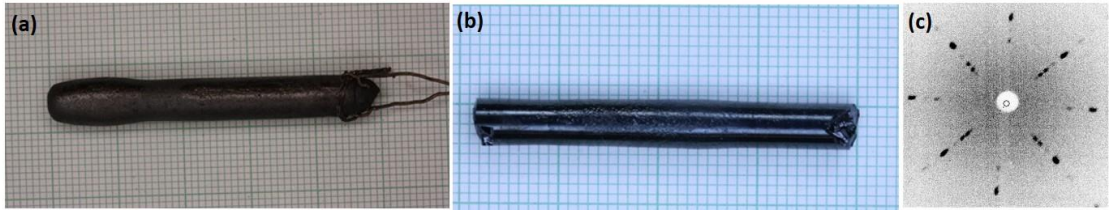

**Figure 8** Photographs of the Nd-LSCO ($x = 0.19$) samples: **(a)** the ceramic rod used for the crystal growth, **(b)** the as-grown single crystal from the OFZ furnace growth, **(c)** Laue diffraction pattern performed on surface of the as-grown single crystal, along the growth direction, [001]. The black spots mark the intensity from the backscattered Bragg peaks, while the white circle shows the centre of the beam.

## *2. Elastic neutron scattering from large single crystals of Nd-LSCO*

While X-ray diffraction can be very useful in refining powder structures, examining the crystallography of small single crystals, and for orienting known single crystals, the structure of large volumes of materials such as Nd-LSCO, either single crystals or otherwise, must be carried out with neutron scattering, so that the full volume of the growth can be sampled and characterized. We carried out elastic neutron scattering measurements on three single crystals of Nd-LSCO, $x = 0.12, 0.19$, and $0.24$, resulting from these growths. These measurements were part of a larger program of elastic and inelastic neutron scattering characterization of these materials. In what follows, we describe only the elastic neutron scattering and how it informs on the single crystal nature of the large samples grown.

**Figure 9** shows maps of the elastic neutron scattering ($-4$ meV $< E < +4$ meV) within the (H, 0, L) plane of reciprocal space taken at $T = 5$ K. There is also an integration required in **Q** space to produce such a two-dimensional map, and this is an integration in K of the form (0, K, 0) from $-0.75 < K < +0.75$. Such maps should only show Bragg intensity at integer positions within the (H, 0, L) plane presented in this way, if the sample is a single crystal, and that indeed is what we observed for the large single crystals of Nd-LSCO $x = 0.12$ in **Figure 9a**, $x = 0.19$ in **Figure 9b** and $x = 0.24$ in **Figure 9c**. Note that the rings of elastic scattering observed in all three data sets are due to Bragg scattering from the aluminium cryostat windows and sample holders, and this is clearly polycrystalline in nature. We therefore conclude that our large single crystals are predominantly single grains of Nd-LSCO.

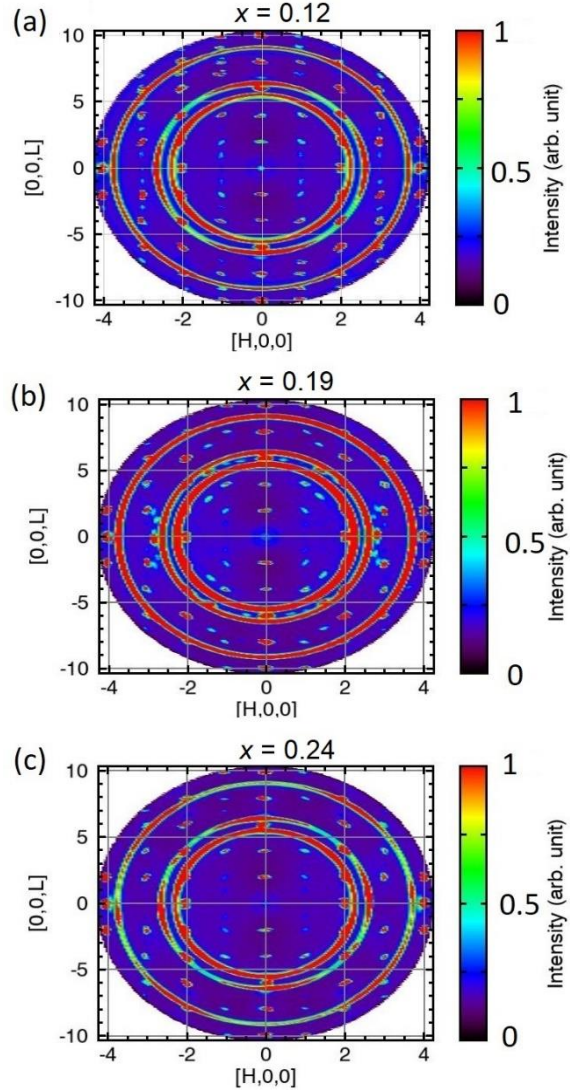

**Figure 9** The elastic Bragg scattering within the (H, 0, L) scattering plane, integrated in K along (0, K, 0) for −0.75 < K < 0.75, and in energy for −4 meV < $E$ < +4 meV are shown at $T$ = 5 K for the three single crystals of $x$ = 0.12 (**a**), 0.19 (**b**) and 0.24 (**c**). The data was collected using the SEQUOIA time-of-flight direct geometry chopper spectrometer. It can be seen that the Bragg peak positions align well with integer positions. The rings of scattering are due to Debye-Scherrer cones of elastic scattering from polycrystalline Al in the windows of the low temperature cryostats and the sample holders.

## 3. High resolution single crystal X-ray diffraction

The structural phase transitions associated with the single crystals of Nd-LSCO which we grew were studied using high-resolution X-ray diffraction. High resolution is required to characterize the small splitting of several of the HTT Bragg peaks once the materials enter orthorhombic structures, such as the LTO and LTT structures. This experiment employed a four-circle X-ray diffractometer and relatively tight collimation to produce the high angular or **Q**-resolution. In this case of the data on the $x = 0.24$ single crystal shown in **Figure 10b**, Cu K$_{\alpha1}$ radiation is not completely separated from Cu K$_{\alpha2}$ radiation, and a "ghost" of the diffraction signal from residual Cu K$_{\alpha2}$ radiation can be seen in the experimental results. **Figure 10** shows the temperature dependence of longitudinal scans through the $(330)_{HTT}$ Bragg peak for single crystals with $x = 0.12$, 0.24 and 0.26, for the 40–200 K temperature interval.

From the neutron powder diffraction experiments we found that at room temperature, the $x = 0.12$ sample adopts the orthorhombic LTO1 structure. Our single crystal X-ray diffraction studies of this sample in **Figure 10a** show that it undergoes a transition to the LTO2 structure at $T \sim 68(4)$ K.

The $x = 0.17$ sample (not shown here) is orthorhombic at room temperature, with the LTO1 structure and a much larger orthorhombic splitting of the $(330)_{HTT}$ Bragg peak than for the $x = 0.24$ sample shown **Figure 10b**. A first order phase transition to the LTT phase was observed at 75(3) K for the $x = 0.17$, sample, only slightly above that displayed by the 0.24 single crystal. As **Figure 10b** illustrates, at room temperature, the $x = 0.24$ sample adopts the tetragonal HTT structure with the $I4/mmm$ space group. The $(330)_{HTT}$ peak continuously splits into (600) and (060) below 150(3) K, as the crystal structure lowers its symmetry to orthorhombic LTO1 (*Bmab* space group) due to the tilting or rotation of the CuO$_6$ octahedra. A second phase transition to the tetragonal LTT ($P4_2/ncmz$) structure occurs at a temperature of about 65(5) K, as is evidenced by a change in the splitting of this set of Bragg peaks. Note that while the longitudinal scans show a narrower profile for the LTT phase than for the LTO1 phase, the LTT phase profile remains clearly broadened compared to that in the HTT phase, consistent with the fact that the LTT phase is fact orthorhombic.

**Figure 10c** shows very similar longitudinal scans for the $x = 0.26$ single crystal sample, and at this concentration no evidence for any orthorhombic splitting is observed at any temperature between 300 K and ~ 4 K. We therefore conclude that the structure of the $x = 0.26$ sample is HTT at all temperatures above 4 K.

The structural transition temperatures identified in our studies are all in excellent agreement with those reported earlier by Axe *et al.* [9]. In particular, our values of $T_1 \sim 150$ K for the HTT to LTO1 transition and $T_2 \sim 60$ K for the LTO1 to LTT transition at $x = 0.24$, are nicely consistent with their values of $T_1 \sim 130$ K and $T_2 \sim 50$ K at $x = 0.25$. Combining our data with theirs, we arrive at the following value for the critical doping at which the LTT phase ends, at $T = 0$: $p_{LTT} = 0.255 \pm 0.005$ (and also consistent with our polycrystal data above; **Figure 6**).

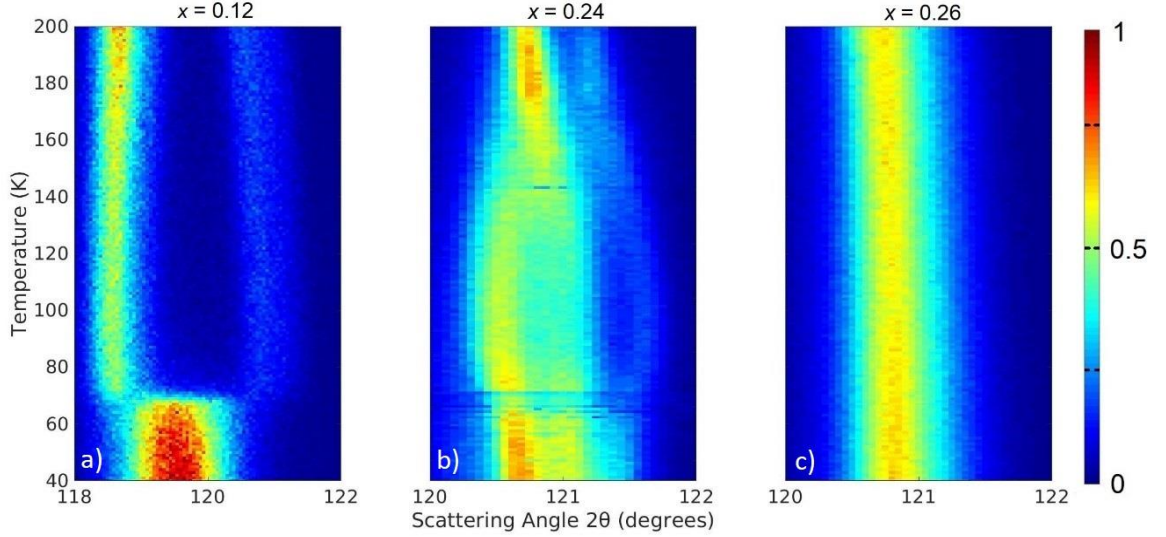

**Figure 10** The temperature dependence of $(3,3,0)_{HTT}$ Bragg peaks monitored via longitudinal X-ray scattering scans for representative Nd-LSCO single crystals with **(a)** $x = 0.12$, **(b)** $x = 0.24$, and **(c)** $x = 0.26$.

A summary of the results of the high-resolution X-ray scattering study of single crystals of Nd-LSCO are presented in **Table III** below.

**Table III** Summary of the structural transition temperatures as derived from X-ray single crystal diffraction measurements performed on the Nd-LSCO single crystals.

| $x$, Sr | 0.12 | 0.17 | | 0.19 | | 0.24 | | 0.26 |
|---|---|---|---|---|---|---|---|---|
| $T_1$   $T_2$ (K) | 68(4) | 290(2) | 75(3) | 275(5) | 70(2) | 150(3) | 65(5) | < 4 K |
| Transition | LTO1 to LTO2 | HTT to LTO1 | LTO1 to LTT | HTT to LTO1 | LTO1 to LTT | HTT to LTO1 | LTO1 to LTT | |

## C. Characterization of the superconducting transition temperatures

A key feature of the high $T_C$ phase diagram is the hole-doping $p$, or $x$, dependence of the superconducting transition temperatures across the Nd-LSCO series. These have been estimated by magnetic susceptibility measurements of our single crystal samples. The susceptibility measurements were performed with the magnetic field applied along the $c$ direction, that is normal to the $ab$ basal plane of the crystals, and two sets of temperature dependent measurements were performed, following both zero-field cooled (ZFC) and field cooled (FC) protocols. The ZFC data shows an abrupt and large drop off in the susceptibility at superconducting $T_C$. The FC data, which shows the Meissner signal, tends to display a much smaller diamagnetic anomaly, which can nonetheless be useful in obtaining a consistent estimate for $T_C$, especially at high doping. The superconducting $T_C$ can then be determined by the temperature at which the ZFC data splits from the FC data, and develops a large diamagnetic signal, as shown in **Figure 11.**

Portions of the single crystal samples with $x \geq 0.17$ were subjected to annealing in flowing oxygen gas, to optimize their oxygen stoichiometry and the sharpness of the superconducting $T_C$'s, and the data in **Figure 11** for $x \geq 0.17$ are all post-annealing. This can be quite important as the condition that the hole concentration equals the Sr stoichiometry, $p = x$, requires that the oxygen content be stoichiometric. The annealing protocol was as follows: the samples were annealed in an oxygen flow of ~15 mL/min under 1 atmosphere, at 950 °C, for 48 hours. The heating and cooling rate was 100 degrees per hour to prevent thermal shock to the crystal. The annealing protocol slightly increased $T_C$, by $\leq 1$ K and sharpened the superconducting transitions in all the samples. A similar tendency was observed in the studies on the LSCO system by Lorentz *et al.* [37] and Lee *et al*. [38,39].

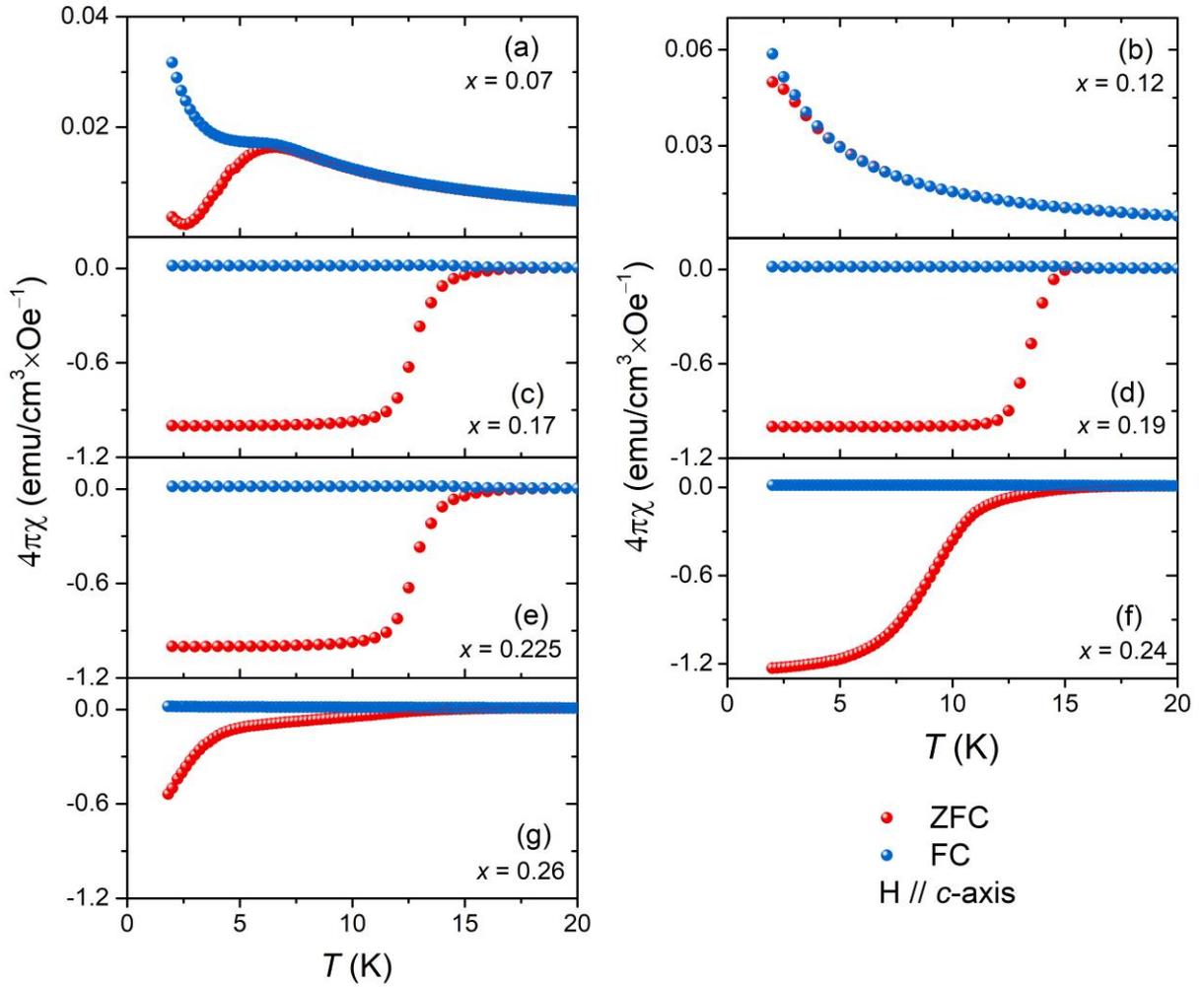

**Figure 11** Magnetic susceptibility data measured on the Nd-LSCO single crystals with $x$ = 0.07, 0.12, 0.17, 0.19, 0.225, 0.24, and 0.26. The latter measurements for $x \geq 0.17$ were performed on sample that had been subjected to an annealing protocol (see text). The magnetic field, $H$, was applied // $c$-axis, $H$ = 10 Oe for 0.17, 0.19, 0.225 and $H$ = 100 Oe for 0.24 and and 0.26 samples.

The superconducting $T_C$ tends to be more rounded at higher doping, and here we employ several experimental measurements in order to provide a robust estimate for $T_C$. **Figure 12a** shows the FC magnetic susceptibility in our Nd-LSCO $x$ = 0.24 single crystal measured with H // $c$ at a range of magnetic field strengths between 10 Oe and 250 Oe, while **Figure 12b** shows the corresponding ZFC magnetic susceptibility taken with $H$ = 100 Oe for $H$ // $c$. We correlate these measurements with a measurement of the in-plane resistivity in zero magnetic field on a piece of the same single crystal, where we see that zero resistance is achieved for 11 K. The FC susceptibilities are all actually paramagnetic (i.e. positive), but clear diamagnetic cusps are observed just at and above 11 K, depending on the strength of the applied magnetic field. The ZFC magnetic susceptibility shows its strongest diamagnetic downturn at 11 K, as well. As a result, we assign $T_C = 11 \pm 1$ K for our $x = 0.24$ Nd-LSCO sample, as marked by the vertical dashed lines in **Figure 12**, consistent with all three of the resistivity as well as FC and ZFC susceptibility measurements (and with previous data [15]).

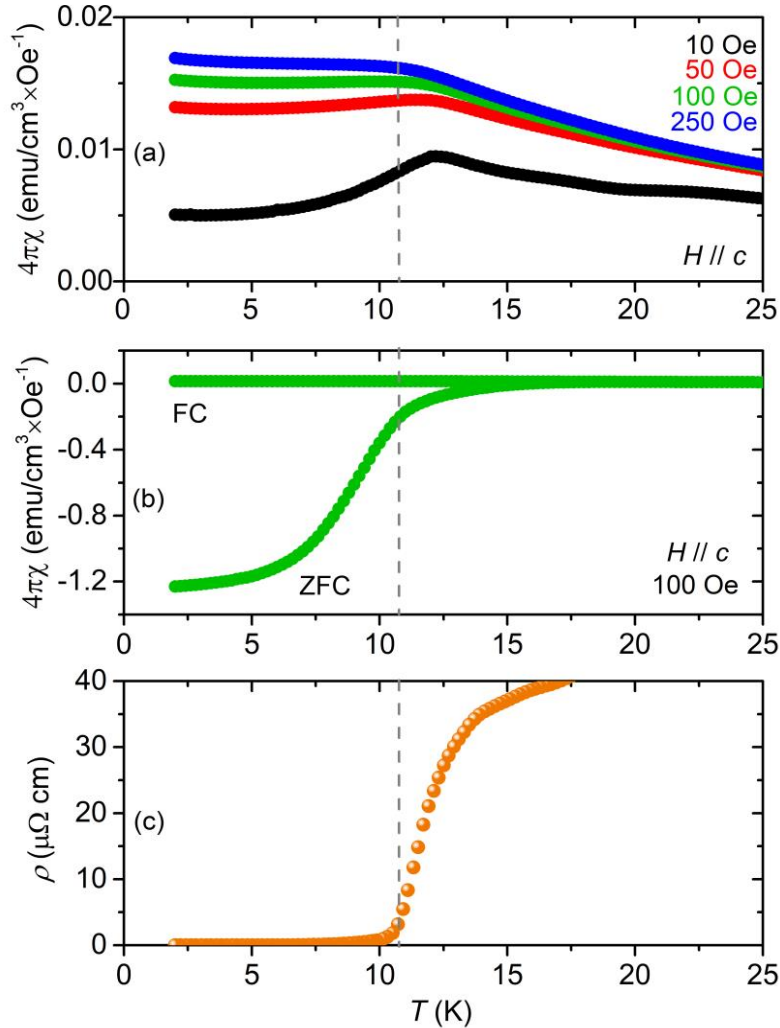

**Figure 12 (a)** FC magnetic susceptibility data on $x = 0.24$ single crystal with the magnetic field applied parallel to $c$-axis for 10 Oe, 50 Oe, 100 Oe and 250 Oe. **(b)** ZFC-FC magnetic susceptibility data on the $x = 0.24$ sample with a field of 100 Oe applied parallel to $c$-axis. **(c)** In-plane resistivity measurements on $x = 0.24$ sample. All data are consistent with a bulk superconducting $T_C = 11 \pm 1$ K.

### D. The superconducting phase diagram in La$_{1.6-x}$Nd$_{0.4}$Sr$_x$CuO$_4$

The superconducting $T_C$'s as a function of doping, $x$, for the seven single crystals of Nd-LSCO we have grown and characterized are plotted in **Figure 13**, along with those determined for single-crystal samples of Nd-LSCO (from Michon *et al*. [15] for $0.12 \le x \le 0.25$) and for polycrystalline samples of Nd-LSCO (from Axe & Crawford [9] for $x < 0.12$). One sees very good consistency between the measured $T_C$'s on our large single crystals and those previously determined from both single-crystal and powder samples available in the literature, with a relatively flat $T_C(x) \sim 15$ K between $x \sim 0.15$ and $x \sim 0.22$. As expected, the overall shape of the superconducting phase diagram is similar to other hole-doped cuprates, with a very pronounced 1/8 anomaly at $x = 0.125$.

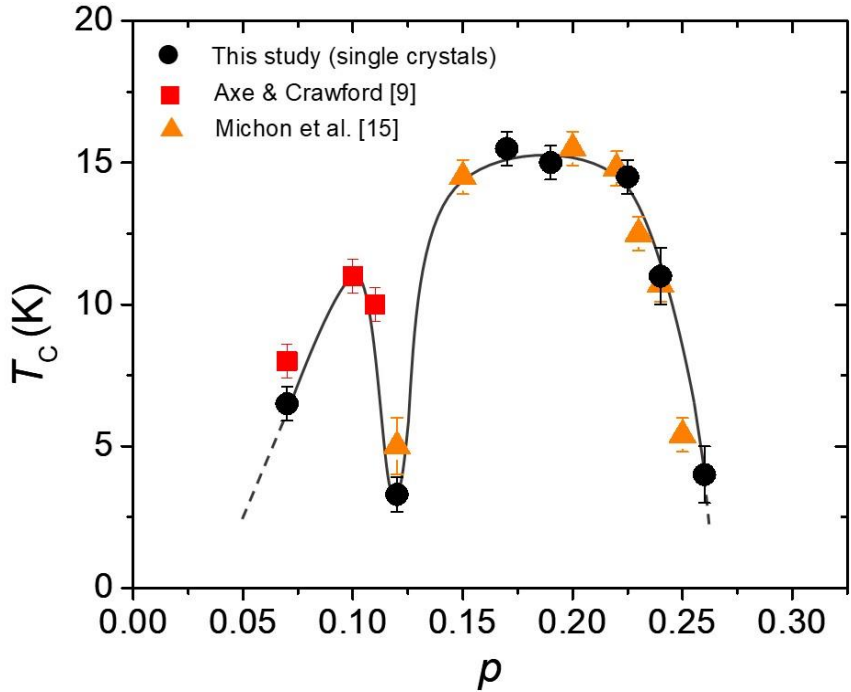

**Figure 13** Superconducting $T_C$ versus $p$ (or $x$) for several sets of Nd-LSCO samples. The present study shows the $T_C$ from the 7 single crystal samples considered (black circles), while the remaining data are from single-crystal samples (orange triangles [15]) and polycrystalline samples (red squares [9]) in the literature.

### E. The structural phase diagram of La$_{1.6-x}$Nd$_{0.4}$Sr$_x$CuO$_4$

The structural phase transitions as measured in the present study with single crystal X-ray diffraction on seven Nd-LSCO samples are summarized in **Figure 14** along with the corresponding transitions determined earlier for polycrystalline samples. This plot therefore summarizes the key results from the present study, and illustrates the level of consistency in the measured phase transition temperatures between our new work on single crystals of Nd-LSCO out to relatively high doping, and earlier measurements on polycrystalline samples. Superposed on the same plot are the measured superconducting $T_C$'s, as shown in **Figure 13**, as well as the values of the pseudogap transition temperature, $T^*$, taken from Cyr-Choinière *et al*. [16], and the associated pseudogap critical point $p^* = 0.23 \pm 0.005$ [15,19,20,22]. We see that $p^*$ is distinctly below the $T = 0$ structural transition at $p_{LTT} = 0.255 \pm 0.005$. This means that one can investigate the effect of the pseudogap phase *within the same structural phase* by comparing Nd-LSCO samples with $p = 0.24$ or $p = 0.25$ on the one hand, to samples with $p = 0.22$ or lower, on the other hand.

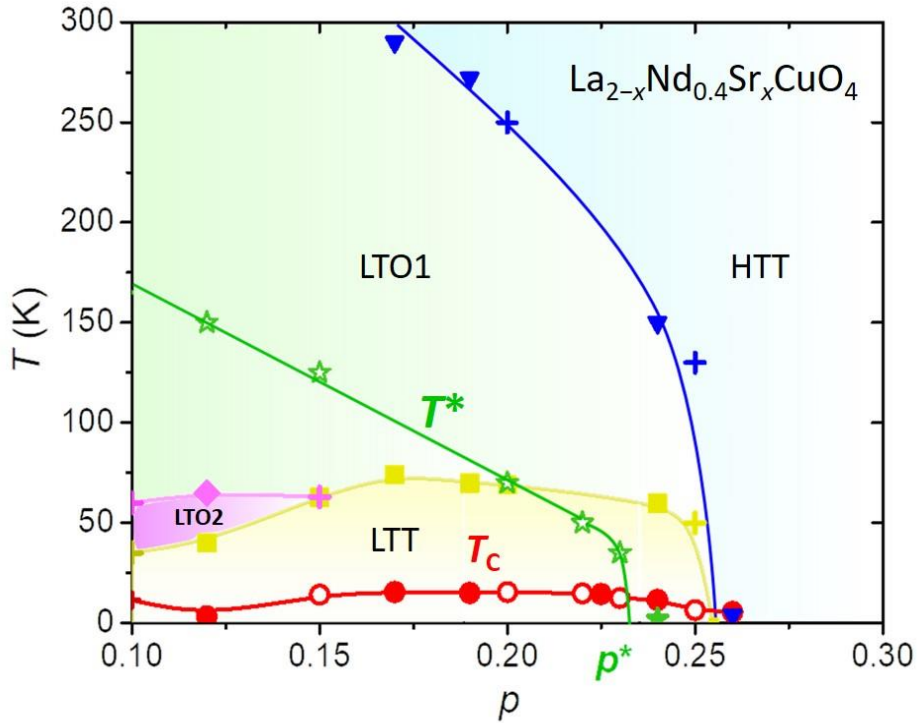

**Figure 14** Proposed Nd-LSCO phase diagram as a function of nominal hole concentration, *p*, or Sr-doping, *x*, and temperature, *T*. The filled symbols are results obtained in this study, the crosses are from Axe & Crawford [9], while the open red circles are reproduced from Michon *et al*. [15]. the pseudogap temperature, $T^*$, is taken from Cyr-Choinière *et al*. [16]. The critical doping $p^*$ where the pseudogap phase ends at $T = 0$ is $p^* = 0.23$ [20].

The corresponding phase diagram for the $La_{2-x}Sr_xCuO_4$ is shown for comparison in **Figure 15**. This plot uses earlier data from Yamada *et al*. [40], and again overplots the line of $T^*$ vs *x* transitions to the pseudogap phase [16]. A striking feature of this plot is that $T^*$ vs *x* is very similar between $La_{1.6-x}Nd_{0.4}Sr_xCuO_4$ and $La_{2-x}Sr_xCuO_4$ for $x \leq 0.15$, but then drops more precipitously to 0 at $p^* \sim 0.18$ in $La_{2-x}Sr_xCuO_4$.

The fact that $p^*$(LSCO) < $p^*$(Nd-LSCO) also correlates with the lower *p* occurrence of the HTT phase at $T = 0$ in LSCO compared to Nd-LSCO, as well as the lower *p* occurrence of the van Hove singularity point ($p^{*}vHs$) in LSCO, which may have some impact on the physics of the pseudogap [41].

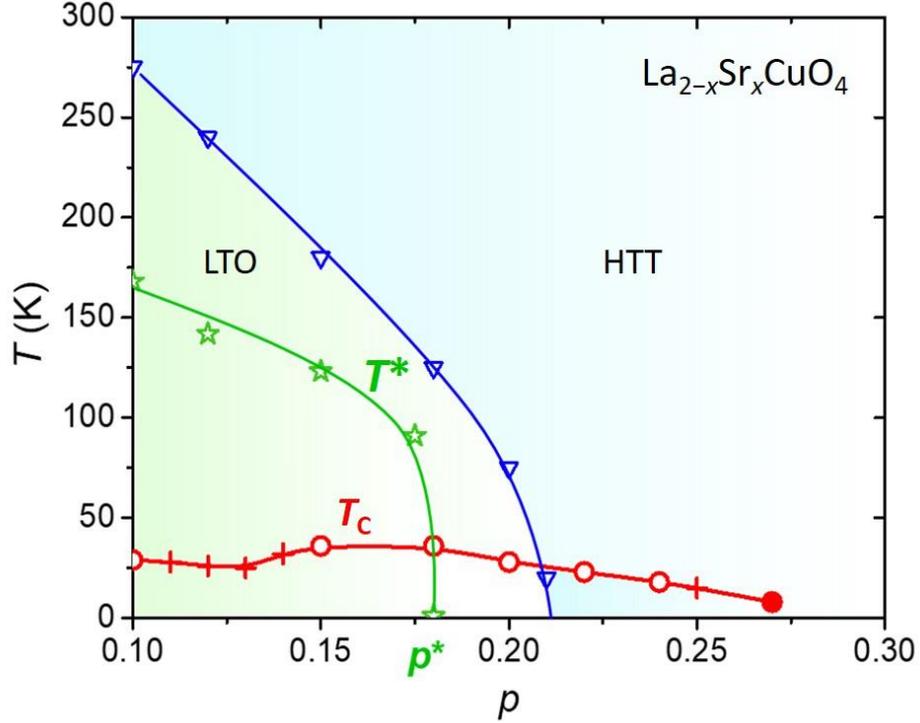

**Figure 15** The LSCO phase diagram, plotted for comparison to that of Nd-LSCO in Figure 14. Empty symbols are from Yamada *et al*. [40], crosses are from Takagi *et al*. [42], filled symbols from Lipscombe *et al*. [43]. The pseudogap temperature, $T^*$, is taken from Cyr-Choinière *et al*. [16]. The critical doping $p^*$ where the pseudogap phase ends at $T = 0$ is $p^* \sim 0.18$ [16].

## IV. CONCLUSIONS

A comprehensive series of $La_{1.6-x}Nd_{0.4}Sr_xCuO_4$ samples, both polycrystalline and single crystals, have been synthesised with $x$ spanning from 0.01 to 0.40, and their structural properties and superconducting $T_C$'s have been investigated by a variety of experimental techniques including X-ray powder and single crystal diffraction, neutron powder diffraction, neutron elastic scattering, as well as by magnetic susceptibility and resistivity measurements. These measurements and their analyses have allowed us to map the structural and superconducting phase transitions that take place as a function of temperature and Sr concentration, $x$, or hole-doping, $p$, and to propose an updated temperature-doping phase diagram for Nd-LSCO, presented in **Figure 14** which now covers both underdoped ($x < 0.05$) and overdoped ($x > 0.25$) regions. This work allows us to compare the phase transitions displayed by new, large single crystals of Nd-LSCO to those from the literature on both single-crystal and polycrystalline samples, and excellent consistency is seen. We conclude that the stoichiometry of our new single crystals is well understood. We find that the critical doping for the end of the LTT structural phase at $T = 0$ is $p_{LTT} = 0.255 \pm 0.005$, above which doping the material appears only in the HTT phase. This means that the onset of the pseudogap phase, at $p^* = 0.23 \pm 0.005$ [15,20], occurs well within the LTT structural phase ($p^* < p_{LTT}$). In other words, the structural and pseudogap critical points are well separated in Nd-LSCO, as they are in LSCO, where $p_{LTO} = 0.215 \pm 0.005$ [40].

# ACKNOWLEDGEMENTS

We wish to thank Jianshi Zhou and Zongyao Li for valuable insights into the crystal growth of Nd-LSCO, as well as Graeme M. Luke and John E. Greedan for discussions and other helpful contributions. This work was supported by the Natural Sciences and Engineering Research Council of Canada.

B.D.G. and L.T. acknowledge support from the Canadian Institute for Advanced Research (CIFAR) as Fellows and from the Canada Foundation for Innovation (CFI). L.T. acknowledges support from the Fonds de recherche du Québec – Nature et Technologies (FRQNT), and the Canada Research Chair program.

We are grateful for the instrument and sample environment support provided during our neutron scattering measurements at the Spallation Neutron Source at Oak Ridge National Laboratory. This part of the research was sponsored by the US Department of Energy, Office of the Basic Energy Sciences, Scientific User Facilities Division.

M. D. and Q. M. contributed equally to this work.